\documentclass[twocolumn,english,aps,jcp,superscriptaddress,showpacs,floatfix]{revtex4}
\usepackage{graphicx}
\usepackage{amsmath,amssymb,amsfonts}
\usepackage{natbib}
\usepackage{babel}

\newcommand{\bra}[1]{\langle #1 |}
\newcommand{\ket}[1]{| #1 \rangle}
\newcommand{\kth}{\mathcal{K}_3}
\newcommand{\ko}{\mathcal{K}_1}

\newcommand{\den}{\hat{\rho}_b^{eq}}
\newcommand{\denb}{\rho_{b,W}^{eq}}

\begin{document}

\title{Accurate nonadiabatic quantum dynamics on the cheap: \\making the most of mean field theory with master equations}
\author{Aaron Kelly}\affiliation{Department of Chemistry, Stanford University, Stanford, CA 94305, USA}
\author{Nora Brackbill}\affiliation{Department of Physics, Stanford University, Stanford, CA 94305, USA}
\author{Thomas E. Markland}\email{tmarkland@stanford.edu}\affiliation{Department of Chemistry, Stanford University, Stanford, CA 94305, USA}

\date{\today{}}

\begin{abstract}
In this article we show how Ehrenfest mean field theory can be made both a more accurate and efficient method to treat nonadiabatic quantum dynamics by combining it with the generalized quantum master equation framework. The resulting mean field generalized quantum master equation (MF-GQME) approach is a non-perturbative and non-Markovian theory to treat open quantum systems without any restrictions on the form of the Hamiltonian that it can be applied to. By studying relaxation dynamics in a wide range of  dynamical regimes, typical of charge and energy transfer, we show that MF-GQME provides a much higher accuracy than a direct application of mean field theory. In addition, these increases in accuracy are accompanied by computational speed-ups of between one and two orders of magnitude that become larger as the system becomes more nonadiabatic. This combination of quantum-classical theory and master equation techniques thus makes it possible to obtain the accuracy of much more computationally expensive approaches at a cost lower than even mean field dynamics, providing the ability to treat the quantum dynamics of atomistic condensed phase systems for long times. 
\end{abstract}

\maketitle

\section{Introduction} \label{sec:intro}
The exact treatment of real time nonadiabatic quantum dynamics in condensed phase chemical systems remains a significant challenge that spurs the ongoing development of approximate methods that are accurate, efficient, and can treat systems with a wide range of different forms of interactions. In particular quantum-classical (semiclassical) trajectory based methods offer a hierarchy of approaches, derived from the exact real time path integral formulation of quantum mechanics, that offer different balances between accuracy and computational cost.

At the lowest tier of this hierarchy is Ehrenfest mean field theory (MFT), which neglects all dynamical correlations between between the quantum (subsystem) and classical (bath) degrees of freedom \cite{mft}. Above this lie linearized path integral approaches such as LSC-IVR \cite{lscivr1,lscivr2}, FK-LPI \cite{lscivr3}, PBME \cite{pbme}, and LAND-map \cite{landmap} which, although still mean field in nature, capture some correlation. These methods offer higher accuracy than mean field theory, at the expense of at least an order of magnitude more computational effort arising from the need to average over the mapping variables associated with the quantum subsystem degrees of freedom. Recently it has been shown that partially linearizing the propagator for the electronic degrees of freedom, giving rise to the partially linearized density matrix (PLDM) \cite{pldm} approach and the forward-backward trajectory solution (FBTS) to the quantum-classical Liouville equation \cite{fbts1}, allows more dynamical correlation to be included at relatively little additional cost to fully linearized methods. Introducing further dynamical correlations between the subsystem and the bath adds further accuracy, at the expense of assigning weights and phase factors to the trajectories  \cite{sstp,mj,ildm}. This in turn adds many orders of magnitude to the number of trajectories, which grows rapidly with system dimensionality and time, that must be generated in order to obtain converged properties. 

Since moving up this hierarchy requires orders of magnitude more computational effort, only the lowest tiers are likely to be practical, both now and in the foreseeable future, for nonadiabatic problems containing large quantum subsystems or where on-the-fly treatment of the electronic states is required. At present, this means that one is typically limited to using MFT, or Tully's fewest switches surface hopping (FSSH) algorithm and its variants \cite{tully1,tully2,bittner_rossky,afssh,tully3}, or linearized path integral approaches \cite{miller, thoss_wang, kapral06}. 

The generalized quantum master equation (GQME) offers an alternative way of exactly describing the nonadiabatic evolution of a quantum subsystem by formally recasting the effects of the environment into a memory kernel \cite{gqme1, gqme2}. In the condensed phase the environment comprises of a large number of modes with a broad range of frequencies that couple to the quantum subsystem, leading to a memory kernel that typically decays much more rapidly than the population relaxation time of the subsystem. This separation in time-scales becomes more pronounced as the system-bath coupling strength or the nonadiabaticity is increased. In such regimes, trajectory based approaches suffer from the rapid accumulation errors in lower tier methods and a rapid rise in the computational cost with time in higher tiered approaches. Hence, using trajectory based approaches to calculate the memory kernel, which is short lived compared to the subsystem relaxation time, and then using the GQME to generate the subsystem dynamics, offers massive advantages in terms of both accuracy and computational cost compared to a direct application of the trajectory based approach. Such a realization has previously been shown to be highly effective in the case of higher-tier trajectory based methods \cite{sng2,sng3,mjgqme}. However, it also allows one to make the most of the lower tier approaches by allowing a more accurate treatment of strongly nonadiabatic and coupled problems while retaining their existing strengths in treating weakly coupled and adiabatic regimes.

Here we show that MFT can be combined with the GQME formalism to yield a method that is as accurate as higher tiered trajectory-based techniques such as FBTS and PLDM, but requires a much lower computational effort than MFT alone, without being limited to any particular form of the Hamiltonian. We show how to calculate the memory kernel of the GQME using dynamical trajectories obtained by solving the mean field equations of motion for the system, and how to use these mean field kernels to subsequently generate the reduced dynamics of the quantum subsystem. Finally we demonstrate that our method is more computationally efficient and more physically accurate for treating charge and energy transfer regimes of the spin-boson model than a direct application of MFT.  

\section{Theory } \label{sec:theory}
\subsection{Ehrenfest Mean Field Theory }
A particularly simple and instructive route to derive the Ehrenfest mean field equations of motion is to begin with the quantum-classical Liouville equation \cite{qcle} and neglect correlations in the system-bath dynamics. One begins by considering a system in which only a small subset of the degrees of freedom behave quantum mechanically, and are of interest. This set of degrees of freedom is denoted as the quantum subsystem (or simply as the subsystem), and the remainder of the system is referred to as the bath, and is assumed to behave essentially classically. 

The total Hamiltonian for the entire system is written as a sum of subsystem, bath, and coupling terms,
\begin{equation} \label{eq:H_tot}
\hat{H} = \hat{H}_s + \hat{H}_b + \hat{H}_{sb},
\end{equation} 
where the subscripts $s, b,$ and $sb$ refer to the subsystem, the bath, and the system-bath coupling,  respectively. The time-evolution of the reduced density matrix of the subsystem is defined as
\begin{equation}\label{eq:rdm}
\hat{\rho}_{s} (t) = Tr_b ( \hat{\rho} (t) ) = \int dX \hat{\rho}_W (X,t),
\end{equation}
where $\hat{\rho}$ is the density operator for the entire system and $\hat{\rho}_W$ is its Wigner transform,  $Tr_b$ indicates the partial trace taken 
over the bath degrees of freedom, $X=(R,P)=(R_1,R_2,...,R_{N_b},P_1,P_2,...,P_{N_b})$, and $N_b$ is the number of bath degrees of freedom.  

The quantum-classical Liouville equation  \cite{qcle},
\begin{eqnarray} \label{eq:qcle}
  && \frac{\partial }{\partial t}\hat{\rho}_W (X, t) =-i {\mathcal L}\hat{\rho}_W (X,t),
\end{eqnarray}
describes the time evolution of the density matrix $\hat{\rho}_W(X,t)$, which is a quantum mechanical operator that depends on the classical phase space variables. The quantum-classical Liouville (QCL) operator is
\begin{equation}\label{eq:qcl_op}
i{\mathcal L} \cdot = \frac{i}{\hbar}[\hat{H}_W,\cdot] - \frac{1}{2}(\{\hat{H}_W,\cdot\}
-\{\cdot,\hat{H}_W\}),
\end{equation}
where $[\cdot,\cdot]$ is the commutator, and $\{\cdot,\cdot\}$ is the Poisson bracket in the phase space of the environmental variables. The subscript $W$ refers to the partial Wigner transform over the environmental degrees of freedom in the system. The partial Wigner transform of the density operator, $\hat{\rho}$, is
\begin{eqnarray} \label{eq:wigner}
\hat{\rho}_{W}(R,P) = \frac{1}{(2\pi\hbar)^{N_b}}\int dZ e^{i P \cdot Z} \langle R - \frac{Z}{2} | \hat{\rho} | R +\frac{Z}{2}\rangle.  
\end{eqnarray}

In order to arrive at MFT, one makes the approximation that the density of the system can be written as a product of the subsystem and bath reduced densities at all times. This asserts that there are no dynamical correlations between the subsystem and the bath, as the total density remains in a product state.
\begin{equation}
\hat{\rho}_W(X,t)=\hat{\rho}_s(t) \rho_{b}(X,t),
\end{equation} 
where the bath RDM is $\rho_b(X, t) = Tr_s ( \hat{\rho}_W (X,t))$. Using the approximations in Eqs. (3) and (6),
one obtains the Ehrenfest mean-field equations of motion \cite{chapter}:
\begin{eqnarray}\nonumber
\frac{\partial \hat{\rho}_s(t)}{\partial t} &=&-\frac{i}{\hbar} \left[\hat{H}_s+\left(\int dX \hat{H}_{sb}(R(t)) \rho_b(X,t)\right),\hat{\rho}_s(t)\right],\\
\frac{dR(t)}{dt}&=&\frac{P(t)}{M},\\\nonumber
\frac{dP}{dt} &=&-\frac{\partial (H_{b,W}(X(t))+Tr_s\{\hat{H}_{sb}(R(t))\hat{\rho}_s(t)\})}{\partial R(t)}.
\end{eqnarray} 
The mean field approximation for a subsystem observable, $\hat{O}_s(t)$, can be written as
\begin{eqnarray} 
\langle \hat{O}_s(t) \rangle = \int dX \left(Tr_s \{ \hat{O}_s \hat{\rho}_s(t) \}\right) \rho_{b} (X,t).
\end{eqnarray}  MFT is a highly efficient approximation to quantum dynamics which, while obtaining accurate results in some regimes, fails when quantum effects in the bath are important and in regimes with a nonzero subsystem energy bias.

\subsection{The Generalized Quantum Master Equation}
\label{subsec:GQME}
An alternative approach to formulating the expectation value of a subsystem observable in terms of the density matrix for the full system, is to treat the reduced dynamics of the subsystem via projection operator techniques. One begins with the exact quantum evolution of the full system, governed by the Liouville - von Neumann equation, 
\begin{eqnarray} \label{eq:lvn}
\frac{\partial }{\partial t}\hat{\rho} (t) = -\frac{i}{\hbar}[\hat{H},\rho(t)],
\end{eqnarray}
and focuses only on the evolution of the subsystem degrees of freedom by projecting out the degrees  of freedom of the bath. This can be accomplished by applying a projection operator, $\mathcal{P}$,
\begin{equation}
\mathcal{P} = \hat{\rho}_b^{eq}\otimes Tr_b (\cdot) = \rho_{b,W}^{eq}\int dX  (\cdot),
\end{equation}
and its compliment, $\mathcal{Q} = 1-\mathcal{P}$. 

The form of the coupling part of the Hamiltonian, $\hat{H}_{sb}$ is chosen to be 
\begin{equation}\label{eq:H_sb}
\hat{H}_{sb} = \hat{S} \otimes \hat{\Lambda}, 
\end{equation}
where $\hat{S}$ is a pure subsystem operator, and $\hat{\Lambda}$ is a pure bath operator. We use this factorization for simplicity and, since any coupling Hamiltonian can be written as a sum of such operators, there is no loss of generality  \cite{geva06}. We also write the bath part of the coupling Hamiltonian, $\hat{\Lambda}$, such that its equilibrium thermal average vanishes,
\begin{equation}\label{eq:bath_avg}
\langle \hat{\Lambda} \rangle_{eq} = Tr_b [ \hat{\Lambda} \hat{\rho}^{eq}_b ]  = 0,
\end{equation} 
In the problems that we consider in Sec.~\ref{sec:results} such a condition is naturally satisfied, but in general this condition can be enforced by redefining $\hat{\Lambda}$ relative to its thermal average  \cite{sng1}.

We restrict our considerations to initial states of the Feynman-Vernon type, where the initial density of full system can be factorized in the following manner,
\begin{equation}\label{eq:ini_con}
\hat{\rho}(t=0) = \hat{\rho}_{s}(0) \otimes \hat{\rho}^{eq}_{b},
\end{equation}
where 
\begin{equation}\label{eq:bath_density}
\hat{\rho}^{eq}_{b} = \frac{\exp(-\beta \hat{H}_b)}{Tr_b [ \exp(-\beta \hat{H}_b) ] }
\end{equation} 
is the density operator for the isolated bath in thermal equilibrium.
Under these conditions the exact time evolution of the subsystem RDM is given by the Nakajima-Zwanzig GQME \cite{gqme1,gqme2}
\begin{equation}
\label{eq:GQME}
\frac{d}{dt} \hat{\rho}_s(t)=-i\mathcal{L}_s\hat{\rho}_s(t)-\int^t_0 d\tau \mathcal{K}(\tau) \hat{\rho}_s(t-\tau),
\end{equation}
where the subsystem Liouville operator, $\mathcal{L}_s$, is given by $\mathcal{L}_s = \frac{1}{\hbar}[ \hat{H}_s, \cdot]$, and the memory kernel, $\mathcal{K}$, is given by
\begin{equation}  \label{eq:kern}
\mathcal{K}(\tau)=Tr_b\{\mathcal{L}_{sb} \exp{\left(-i \mathcal{Q} \mathcal{L} \tau\right)} \mathcal{Q} \mathcal{L}_{sb} \den\}
\end{equation}

The assumption of an initially uncorrelated state is not necessary, and relaxing it would introduce an inhomogeneous term to the GQME which depends on the chosen initial state (that can also be calculated by mean field theory and other approximate methods). The free subsystem evolution prescribed by $\mathcal{L}_s$ is generally simple to simulate, and hence calculating the evolution of the subsystem RDM reduces to the calculation of the memory kernel, which encodes the deviation from free subsystem evolution. From Eq. (15), it is clear that changes in the subsystem populations in the GQME are driven by the subsystem Liouville operator as well as the memory kernel and thus the lifetime of the population dynamics is typically longer than that of the memory kernel. In condensed phase systems, where the environment spans a wide spectral bandwidth, the memory kernel is expected to decay much more quickly than the population relaxation time.

The general form for the memory kernel, given above, is not straightforward to evaluate since it explicitly depends on the projection operator. However, there are a variety of ways that it can be constructed from simulations of the unprojected dynamics of the full system. We will limit our discussion of this procedure to a particular method introduced by Shi and Geva, who showed that the full memory kernel, $\mathcal{K}(\tau)$, can be written in terms of a set of partial memory kernels \cite{sng1,sng2,geva06},
\begin{eqnarray} \label{eq:k_tot}
\mathcal{K}(\tau) & = & \mathcal{K}_1(\tau) + i \int_0^{\tau}d\tau' \mathcal{K}_1(\tau - \tau') \mathcal{K}_2(\tau),\\
\mathcal{K}_2(\tau) & = & \mathcal{K}_3(\tau) + i \int_0^{\tau}d\tau' \mathcal{K}_3(\tau - \tau') \mathcal{K}_2(\tau), 
\end{eqnarray} where the partial memory kernels are given by
\begin{eqnarray} \label{eq:k_13}
\mathcal{K}_1(\tau) & = & Tr_b \{ \mathcal{L}_{sb} e^{-i \mathcal{L}\tau}\mathcal{L}_{sb} \hat{\rho}^{eq}_{b} \},\\
\mathcal{K}_3(\tau) & = & Tr_b \{ e^{-i \mathcal{L}\tau}\mathcal{L}_{sb} \hat{\rho}^{eq}_{b} \}.
\end{eqnarray}
By combining Eq. (15), and Eqs. (17 - 20) the subsystem RDM can thus be exactly evolved using projection-free input. Exact numerical evaluations of these expressions have recently been carried out using techniques such as QUAPI, real-time quantum Monte Carlo, and ML-MCTDH  \cite{sng1,muhlbacher08,cohen11,wilner}.

\subsection{Mean Field Evaluation of the Memory Kernel} \label{mf_eval}
In practice, evolving the subsystem RDM using the GQME formalism for an arbitrary system is no less cumbersome than solving Eq. (9). One way to proceed, that is valid in the weak coupling limit, is to factorize the memory kernel into subsystem and bath parts which can be evaluated separately, leading to the well established Bloch-Redfield theory \cite{sng1,redfield}. In this limit $\mathcal{K}_2$ and $\mathcal{K}_3$ vanish, and the memory kernel can be evaluated using equilibrium correlation functions of the isolated bath.  In contrast, here we make no such simplifying assumption and instead evaluate the kernels using MFT.

Since $\ko$ and $\kth$ do not contain any projected input, they can be simulated directly and then used to obtain $\mathcal{K}$ by solving Eqs. (17) and (18). The matrix elements of $\ko$ and $\kth$ are obtained by projecting each quantity onto a basis which spans the subsystem Hilbert space,
\begin{eqnarray} \label{eq:k_1_elements}
\nonumber(\mathcal{K}_1)_{\alpha \alpha' \beta \beta' }(\tau) &=& \Big\langle S_{\alpha \mu'}(\tau) \hat{\Lambda}^{\beta' \alpha'}_{\mu' \mu} (\tau) S_{\mu \beta}(0)  \hat{\Lambda}(0)  \Big\rangle_{eq} \\\nonumber &&-  \Big\langle S_{\mu' \alpha'}(\tau) \hat{\Lambda}^{\beta' \mu'}_{\alpha  \mu} (\tau) S_{\mu \beta}(0)  \hat{\Lambda}(0) \Big\rangle_{eq} \\\nonumber &&+  \Big\langle \hat{\Lambda}(0) S_{\beta' \mu}(0)   \hat{\Lambda}^{\mu \mu'}_{\alpha  \beta} (\tau) S_{\mu' \alpha'}(\tau) \Big\rangle_{eq} \\\nonumber && -\Big\langle  \hat{\Lambda}(0) S_{\beta' \mu}(0) \hat{\Lambda}^{\mu \alpha'}_{\mu'  \beta}(\tau) S_{\alpha \mu'}(\tau)  \Big \rangle_{eq},\\
\end{eqnarray} 
\begin{eqnarray}\label{eq:k_3_elements}
\nonumber (\mathcal{K}_3)_{\alpha \alpha' \beta \beta' }(\tau) &=& \Big\langle (\hat{1}_b)^{\beta' \alpha'}_{\alpha  \mu}(\tau) S_{\mu \beta}(0) \hat{\Lambda}(0)  \Big\rangle_{eq} \\ &&-  \Big\langle S_{\beta' \mu}(0)  \hat{\Lambda}(0) (\hat{1}_b)^{\mu \alpha'}_{\alpha  \beta}(\tau)  \Big\rangle_{eq},
\end{eqnarray} 
where $\alpha, \alpha', \beta,$ and $\beta'$ refer to subsystem states, and $\hat{1}_b$ is the unit operator for the bath. In the above two expressions the Einstein summation convention is used.

The matrix elements of the partial memory kernels, $\mathcal{K}_1$ and $\mathcal{K}_3$, contain correlation functions of the following form, 
\begin{eqnarray} \label{eq:sdbcf1} \nonumber
\langle \hat{S}(0) \hat{\Lambda} (0) \hat{\Gamma}^{\beta' \alpha'}_{ \alpha  \beta}(\tau) \rangle_{eq} &=& Tr \Big( \hat{\rho}_b^{eq} \hat{H}_{sb}| \beta \rangle \langle \beta' |   \\ &&\times e^{i \mathcal{L}\tau/\hbar} \hat{\Gamma} | \alpha' \rangle  \langle \alpha | \Big),
\end{eqnarray}  
where $\hat{\Gamma}$ is a bath operator ($\hat{1}_b$ or $\hat{\Lambda}$). Defining operators, $\mathcal{A} = \hat{H}_{sb} (\hat{1}_b \otimes |\beta\rangle\langle\beta'|)$ and $\mathcal{B}=\hat{1}_b\otimes |\alpha'\rangle\langle\alpha|$ , or $\mathcal{B} = \hat{H}_{sb} (\hat{1}_b \otimes |\alpha'\rangle\langle\alpha|)$, expression (\ref{eq:sdbcf1}) takes on the general form for a quantum time correlation function, 
\begin{eqnarray} \label{eq:sdbcf2}
\langle \hat{S} (0) \hat{\Lambda} (0) \hat{\Gamma}^{\beta' \alpha'}_{ \alpha  \beta}(\tau) \rangle_{eq} =  Tr(\hat{\rho}_b^{eq}\mathcal{A} \mathcal{B}(\tau)).
\end{eqnarray} where the equilibrium density corresponds to that of the isolated bath. 

Working in the coordinate representation of the bath degrees of freedom, and making use of the partial Wigner transform, Eq. (\ref{eq:sdbcf2}) can be rewritten as
\begin{eqnarray} \label{eq:c_ab}
 Tr(\hat{\rho}_b^{eq}\mathcal{A} \mathcal{B}(\tau)) = Tr_s \int dX \left[\hat{\rho}_b^{eq} \mathcal{A}\right]_W(X,0) \mathcal{B}_W(X,\tau).
\end{eqnarray} 

The calculation of the memory kernel then amounts to the evaluation of the above expression, which can be performed by a hybrid Monte Carlo / molecular dynamics algorithm where (i) initial conditions are sampled from $\left[\hat{\rho}_b^{eq} \mathcal{A}\right]_W(X,0)$, and (ii) the system is propagated in time from these initial conditions using MFT to evaluate $ \mathcal{B}_W(X,\tau)$. The full memory kernel can then be constructed, and the subsystem RDM propagated as follows:
\begin{enumerate}
\item Mean-field trajectories are used to obtain the correlation functions necessary to form $\mathcal{K}_1$ and $\mathcal{K}_3$ using Eqs. (21) and (22). See Appendix A for more details on the calculation of these quantities for the model studied here.
 \item $\mathcal{K}_2$ is generated from $\mathcal{K}_3$ by an iterative solution to Eq. (18), using $\mathcal{K}_3$ itself as an initial guess for $\mathcal{K}_2$. This iterative procedure typically converges very quickly, and often requires only a few tens of iterations. 
 \item $\mathcal{K}_1$ and $\mathcal{K}_2$ are used as input to obtain the full memory kernel $\mathcal{K}$ by numerical integration of Eqs. (17) and (18). 
 \item  Using the full memory kernel, the evolution of the subsystem density is generated by direct numerical integration of the GQME using Eq. (\ref{eq:GQME}).
 \end{enumerate} In our calculations the kernel elements were calculated for the specified time $\tau$, and set to zero for all $t > \tau$; no smoothing of the kernel data was performed for $t < \tau$.  Using the MF-GQME approach one can propagate the subsystem RDM for arbitrarily long times using only short-time information obtained from mean field trajectories.

\section{Results and Discussion} \label{sec:results}
In order to assess the accuracy and efficiency of our MF-GQME approach, we performed simulations of the spin-boson model. Despite its apparent simplicity, this system is a prototypical model for the study of quantum transport and relaxation processes in the condensed phase  \cite{spinboson1,spinboson2}, and remains a challenging test to approximate methods. Since it is now possible to generate numerically exact results in many of the parameter regimes of the spin-boson model, it provides an ideal benchmark test case for the accuracy and efficiency of approximate nonadiabatic dynamics approaches. In particular we compare our MF-GQME approach to a direct MFT treatment, as well as to the recently introduced FBTS method \cite{fbts1,fbts2} which has been shown to outperform fully linearized methods at marginal extra computational cost.  

The spin-boson Hamiltonian can be written in the subsystem basis as 
\begin{equation} \label{eq:sb_ham}
\hat{H} = \epsilon \hat{\sigma}_z + \Delta \hat{\sigma}_x + \frac{\hat{P}^2}{2M} + \sum_j\left(\frac{1}{2}M_j\omega_j^2 \hat{R}_j^2 - c_j \hat{R}_j\hat{\sigma}_z\right),
\end{equation} where $\hat{\sigma}_x$ and $\hat{\sigma}_z$ are Pauli spin matrices. This Hamiltonian describes a two level quantum system with energetic bias $2\epsilon$, and electronic coupling (tunneling) matrix element $\Delta$, that is bi-linearly coupled to a bath of independent harmonic oscillators. In the spin-boson model both subsystem states are coupled to the same bath in an anti-correlated fashion. In contrast to problems where the subsystem states are coupled to independent uncorrelated baths, such as the Frenkel exciton model, this form of the coupling presents a more difficult challenge for mean-field type methods to describe \cite{taomiller,kellyrhee,rdmh1,rdmh2,fbts2,pldm1,ildm2}. The greater challenge of treating anti-correlated baths is due to the greater difference between the mean force and those on the individual diabatic surfaces compared to independent, uncorrelated, baths. 

\begin{figure}[t]
\includegraphics[width=0.8\columnwidth,angle=0]{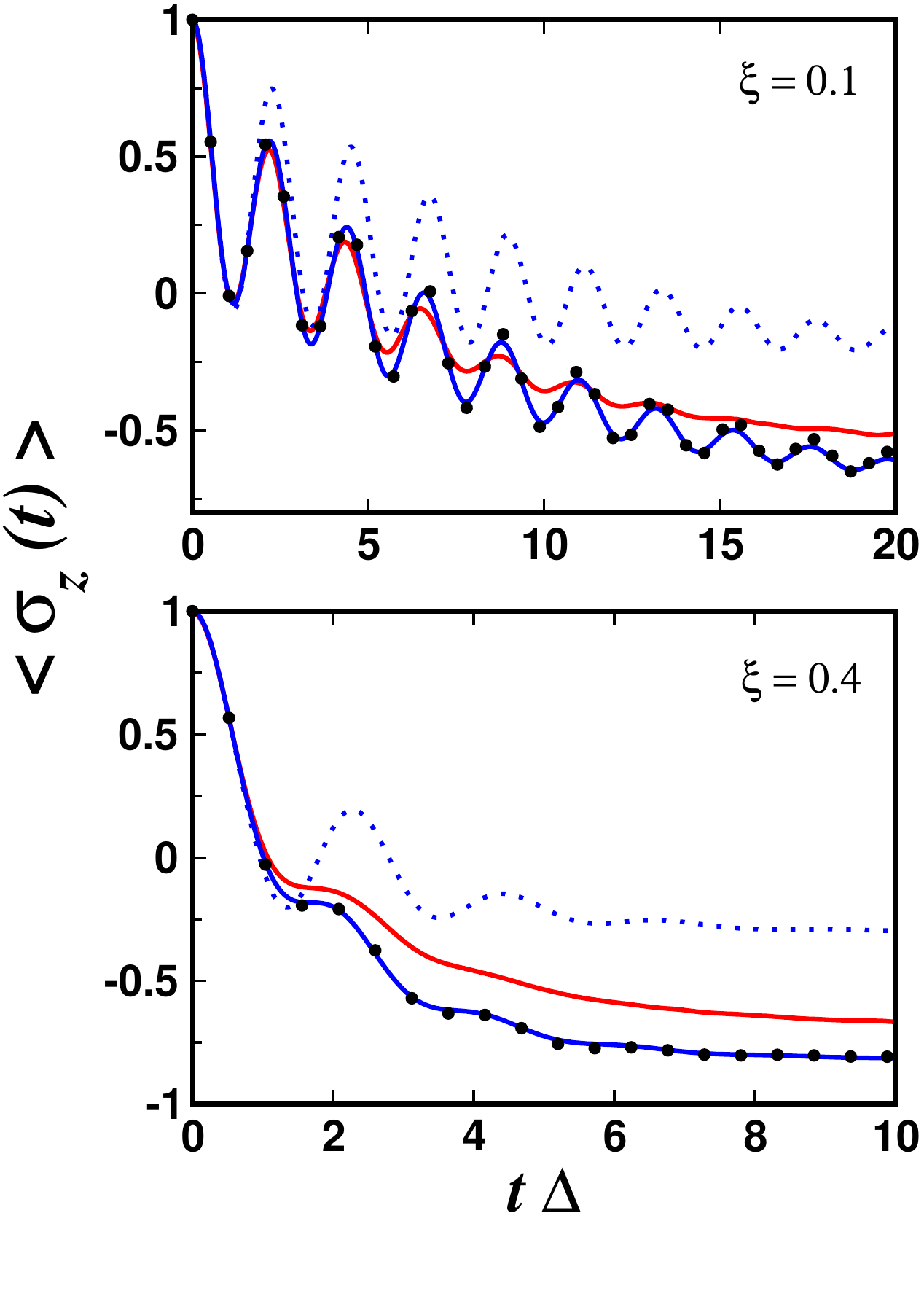} 
\caption{Evolution of the subsystem population difference in the biased, nonadiabatic regime with increasing system-bath coupling strength; $\omega_c= 2 \Delta$, $\epsilon = \Delta$, $\beta = 5 \Delta^{-1}$.  In each panel the exact results are shown in solid black dots, the dotted blue lines are direct MFT results, the solid red lines are FBTS results, and the solid blue lines are the results of our MF-GQME approach.}
\label{fig:2}
\end{figure}
The interaction between the system and the bath can be fully characterized by the spectral density, $J(\omega)$, which determines the strength of the interactions between the subsystem and bath, which we chose to be of Ohmic form, \begin{eqnarray} \label{eq:spec_dens}
J(\omega) & = & \frac{\pi}{2} \xi \omega e^{-\omega / \omega_c}.\end{eqnarray} 
The Kondo parameter, $\xi$, controls the strength of the coupling between subsystem and the bath, and the cutoff frequency $\omega_c$ sets the primary time-scale for the bath evolution. The quantum subsystem was initialized in diabatic state 1, and the bath was sampled from its (isolated) equilibrium distribution. In our calculations 400 bath modes were used to represent the continuous spectral density which, for all regimes and approaches employed, gave results converged to graphical accuracy. 
\begin{figure}[ht]
\includegraphics[width=0.8\columnwidth,angle=0]{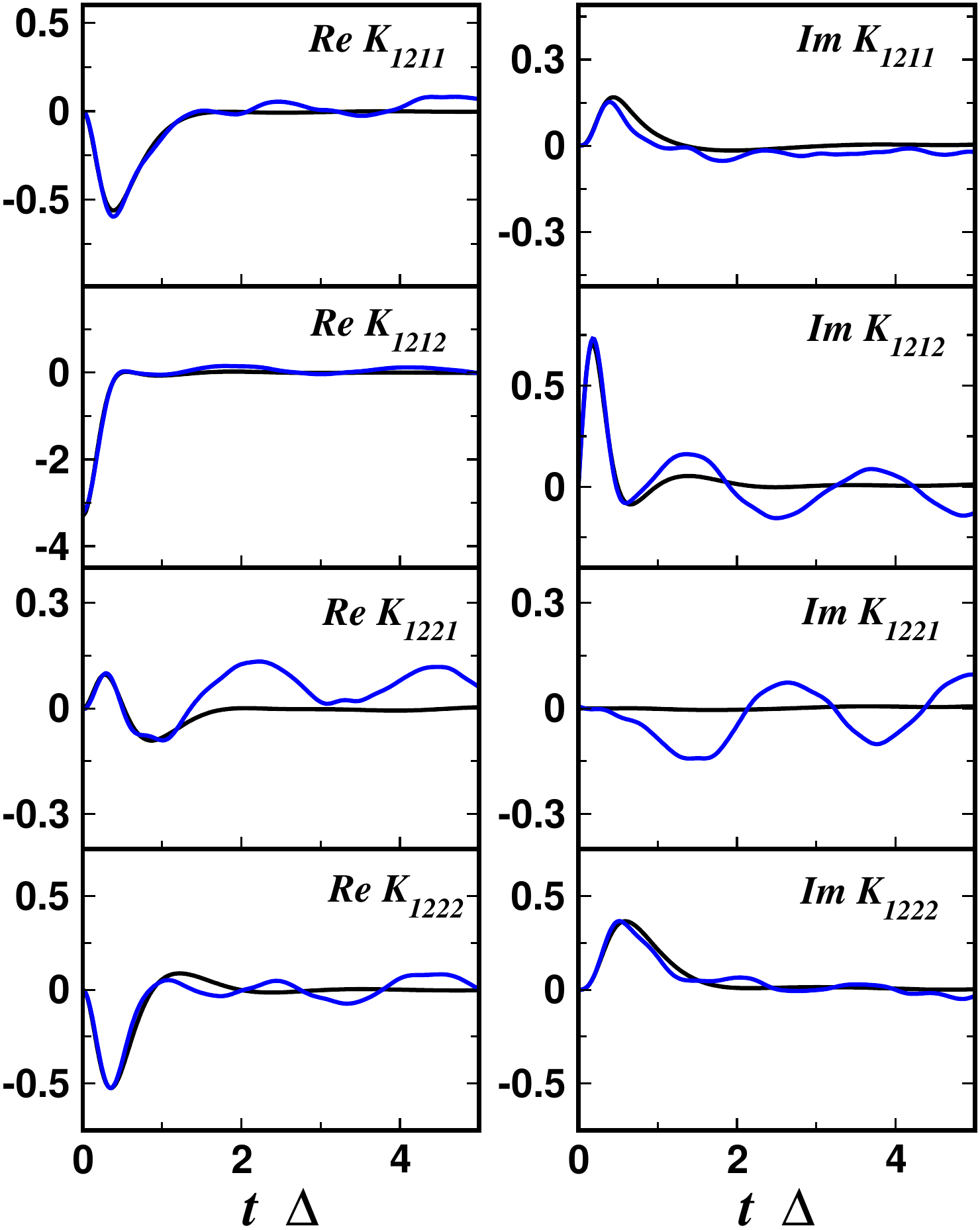} 
\caption{Matrix elements of the memory kernel of the GQME for for $\omega_c= 2 \Delta$, $\epsilon = \Delta$, $\beta = 5 \Delta^{-1}$, $\xi=0.4$, $\delta = 0.02\Delta^{-1}$, and $N_{traj} =2$x$10^4$. In each panel the MFT results are shown as blue lines, and the exact QUAPI results are the black lines. Note the different y-axis scales in each panel, due to the varying magnitudes of each element.}
\label{fig:1}
\end{figure}

\begin{figure}[ht]
\includegraphics[width=0.8\columnwidth,angle=0]{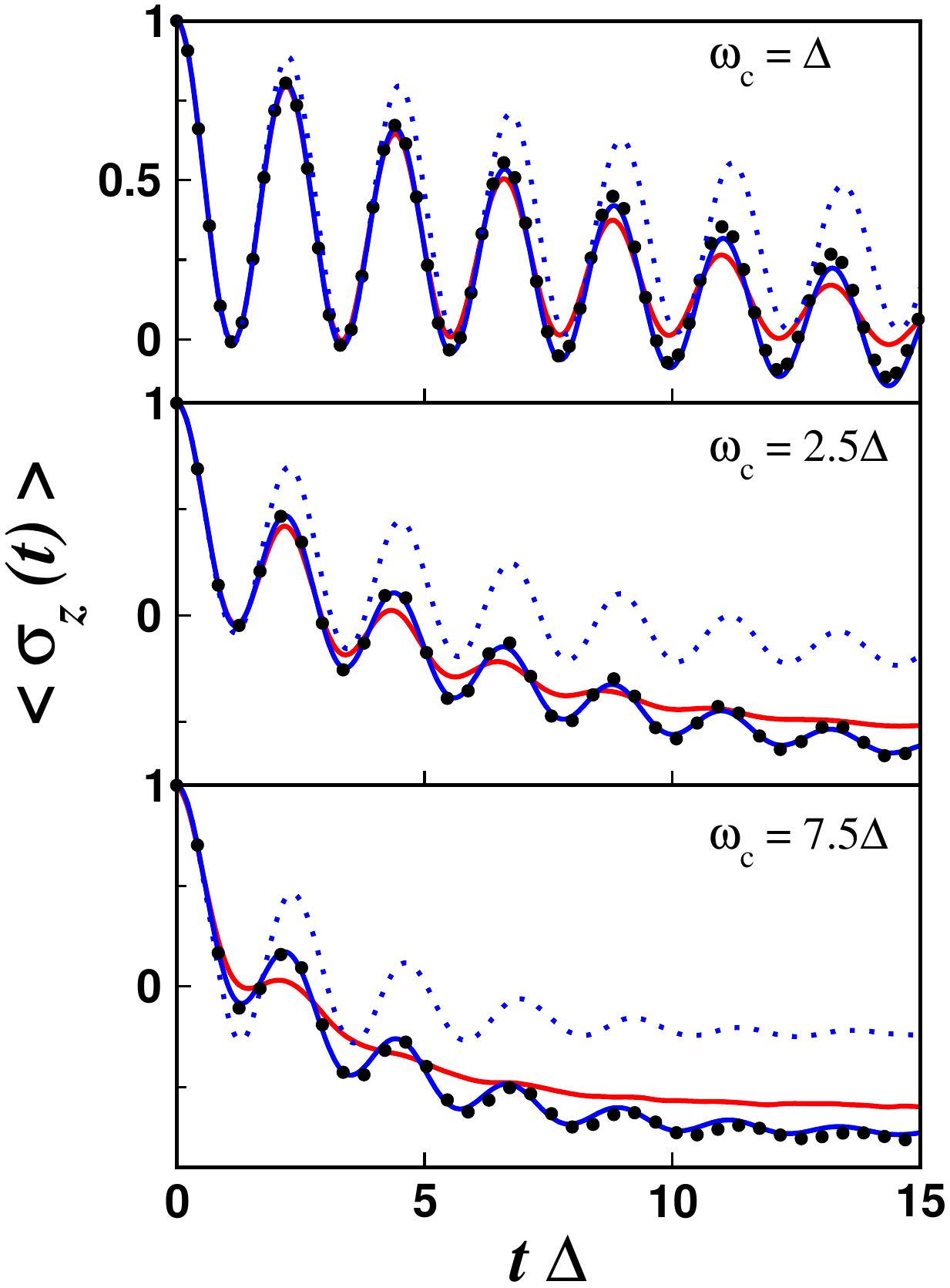} 
\caption{Evolution of the subsystem population difference in the intermediate coupling regime with increasing bath nonadiabaticity; $\epsilon = \Delta$, $\beta = 5 \Delta^{-1}$, $\xi = 0.1$.  In each panel the exact results are the solid black dots, the dotted blue lines are direct MFT results, the solid red lines are FBTS results, and the solid blue lines are the results of our MF-GQME approach.}
\label{fig:3}
\end{figure}
Figure \ref{fig:2} compares the subsystem evolution as the system-bath coupling is increased, for a system with an energetic bias ($\epsilon \neq 0$) in the nonadiabatic regime ($\frac{\omega_c}{\Delta} > 1 $). Numerically exact results were generated using our own implementation of the quasi-adiabatic path-integral (QUAPI) algorithm \cite{quapi1,quapi2,quapi3}. The performance of trajectory based approaches degrades as the system becomes more nonadiabatic and the system-bath coupling is increased, a failure that is particularly evident for systems with nonzero energetic bias. In this regime direct MFT is over-coherent and only captures the exact QUAPI results at very short times ($\tau\Delta < 1.5$). The long time population difference is close to zero, which is a notorious feature of MFT that arises due to the deviation of the mean force from the forces on the individual surfaces, resulting in an accumulation of error in the subsystem population distribution as time progresses. FBTS, which is also mean field in nature but includes more dynamical correlation between the subsystem and bath, much more accurately captures the long time populations at a modest increase in the number of trajectories of by a factor of $\approx 5$ in this regime. In contrast to MFT, the FBTS underestimates the oscillatory nature of the population decay.  

By using MFT to approximate the memory kernel, our MF-GQME approach produces populations dynamics that are in perfect quantitative agreement with the exact quantum results. Figure \ref{fig:1} shows the MFT memory kernel elements that give rise to the dynamics in the bottom panel of Fig. \ref{fig:2}, compared with the exact QUAPI results. There are four nonzero, linearly independent elements of $\mathcal{K}$ for a two-level system (due to the form of the spin-boson Hamiltonian,  $\mathcal{K}_{\alpha \alpha \beta \beta'}  = 0$).  As expected, MFT fails to correctly capture the long time behavior; exhibiting spurious oscillations when $\tau\Delta > 2$. This is consistent with the fact that the direct MFT treatment of the population dynamics in Fig. \ref{fig:2} also begins to show marked errors at those times. Despite the deviations of MFT from the exact kernels, the population dynamics generated using a memory kernel of length $t\Delta=1.5$ (shown in the bottom panel of Fig. \ref{fig:2}) are in excellent agreement with the exact results. Using memory times longer than $\tau\Delta = 1.5$ includes the spurious long time oscillations but only introduces very minor changes to the population dynamics. As MF-GQME only requires the generation of very short trajectories to obtain the memory kernel, the entire population decay, which occurs in a time of approximately 15$\Delta^{-1}$, can be obtained at a cost $10$ times cheaper than a standard mean-field calculation of the same observable and $\approx 50$ times cheaper than FBTS.

The effect of increasing the characteristic frequency of the bath, $\omega_c$, which pushes the system into an increasingly nonadiabatic regime, is displayed in Fig. \ref{fig:3}. In the adiabatic limit MFT is accurate, but as the nonadiabaticity is increased the direct MFT results begin to deviate from the exact solution at progressively shorter times, and by $\omega_c=7.5\Delta$ it is unable to even capture the first minimum in the coherent decay correctly. FBTS again performs better than direct MFT in reproducing the long time limit, albeit at an increased cost of an order of magnitude more trajectories, but again gives results which are increasingly overdamped as $\omega_c$ is increased. This overdamping is consistent with that observed in previous FBTS and PLDM results in similar regimes  \cite{fbts2, pldm1}. 

The MF-GQME results are in quantitative agreement at low nonadiabaticity, and even at the highest nonadiabaticity exhibit only a very subtle phase shift relative to the exact results. Again this reflects that the memory kernel decays rapidly in these nonadiabatic regimes, and hence retaining a memory kernel of length $\tau\Delta = 1.5$ is sufficient to generate the results shown in Fig. \ref{fig:3}.
\begin{figure} 
\includegraphics[width=0.8\columnwidth,angle=0]{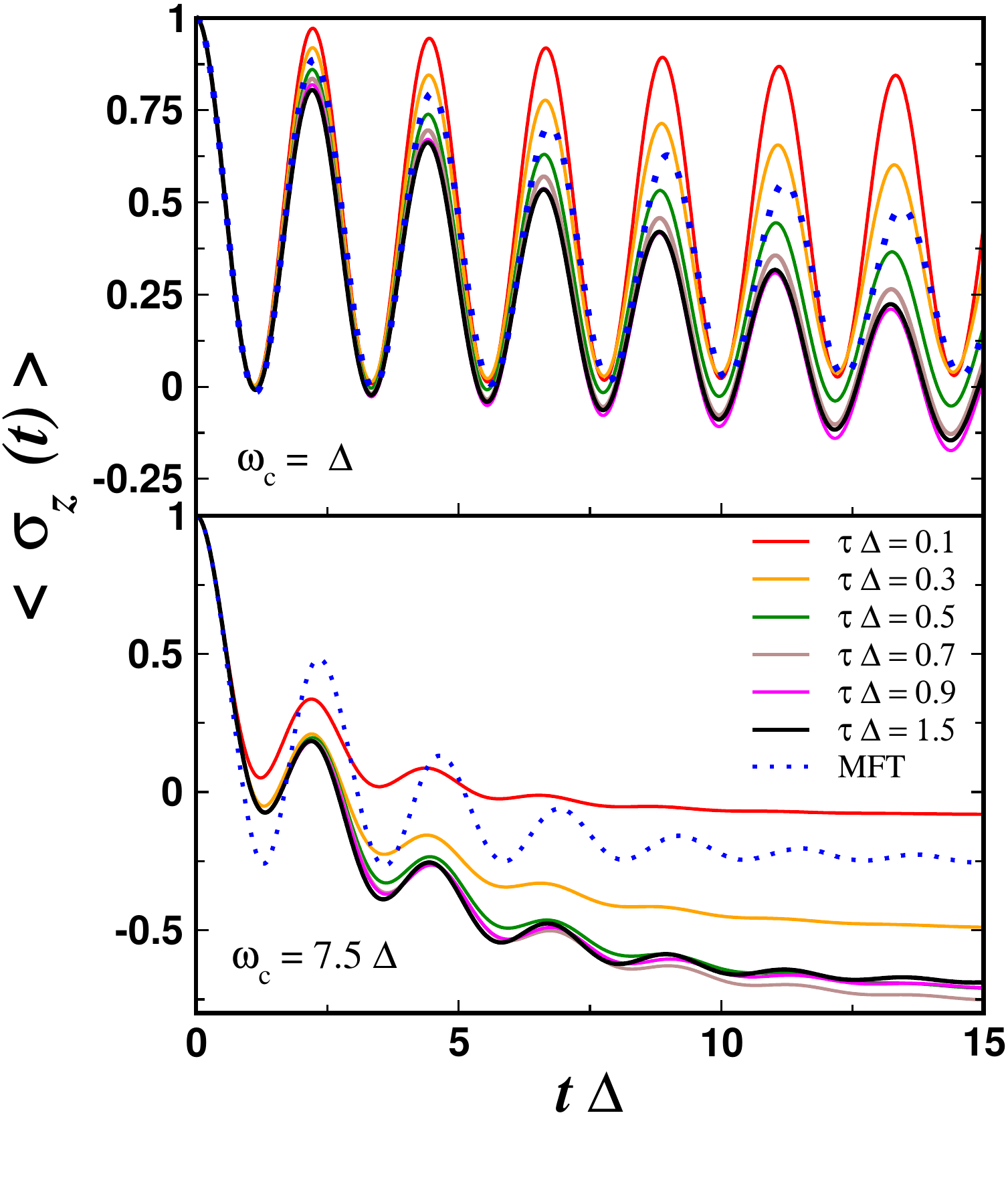} 
\caption{Evolution of the subsystem population difference generated from the MF-GQME approach using memory kernels of varying length; $\epsilon = \Delta$, $\beta = 5 \Delta^{-1}$, $\xi = 0.1$.}
\label{fig:4}
\end{figure}
Figure \ref{fig:4} shows the convergence of the population dynamics obtained from the MF-GQME approach when different amounts of time, $\tau\Delta$, are included in the memory kernel for the regimes shown in the top and bottom panels of Fig. \ref{fig:3}.  In both cases the convergence is essentially monotonic as the length of memory in the kernel is increased. The MF-GQME results are generally better than direct MFT even for very short memory kernels. When an insufficient length of time is used in the kernel, the error accumulated in the propagation of the subsystem RDM in MF-GQME manifests in the observed population dynamics more prominently at longer times.  

For the more adiabatic regime shown (top panel) the dissipation induced by the bath is well captured when $\tau\Delta = 0.7$, whereas when the system becomes more nonadiabatic (bottom panel) this is decreased further with convergence obtained around $\tau\Delta = 0.5$, which is $~30$ times shorter than the population decay time. This again highlights the complementarity of using MFT and other trajectory based approaches in conjunction with the GQME framework; moving further into the nonadiabatic regime, which leads to a faster breakdown of MFT, requires less total time to be included in the GQME memory kernel.

\begin{figure}
\includegraphics[width=0.8\columnwidth,angle=0]{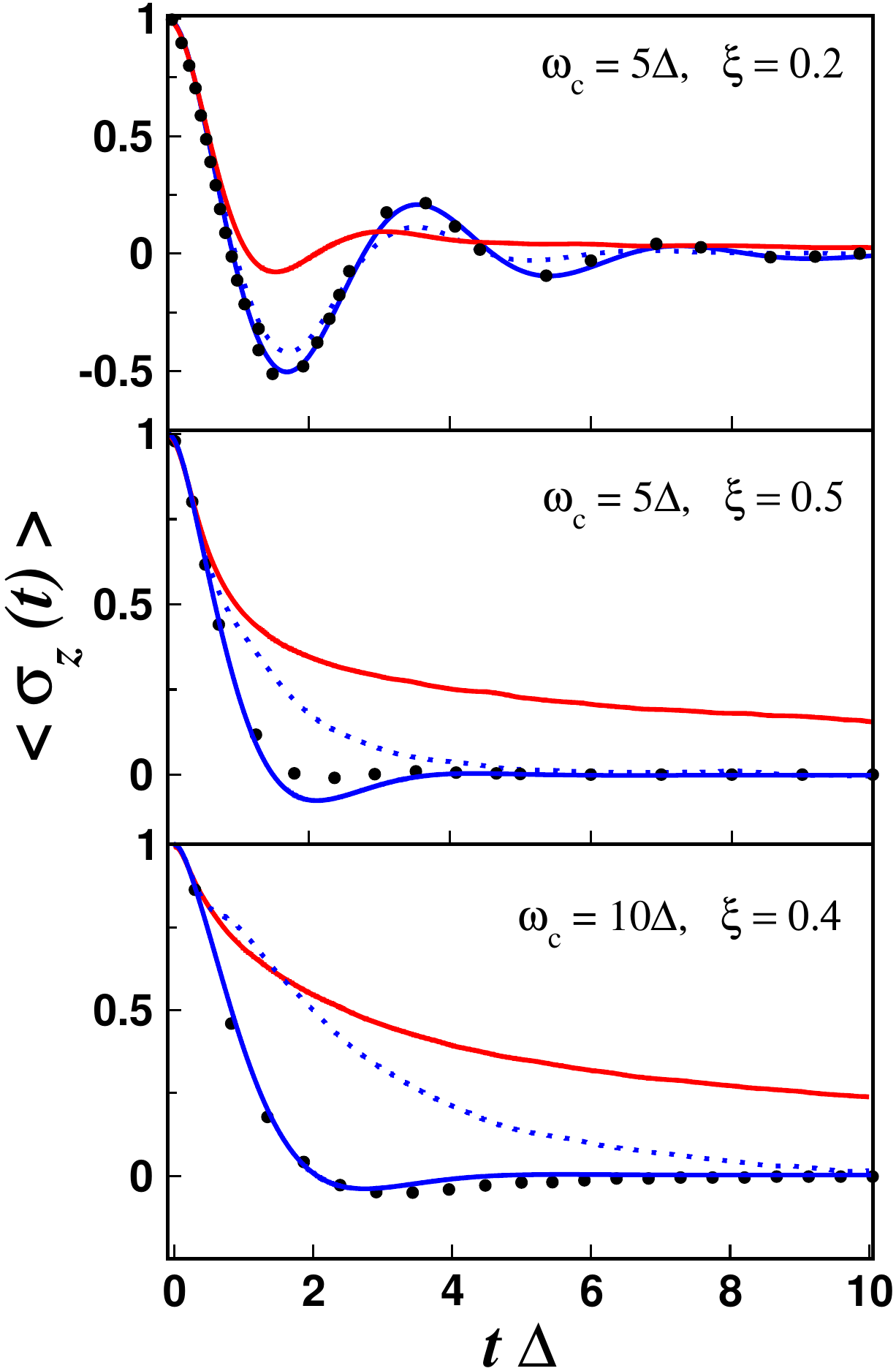} 
\caption{Evolution of the subsystem population difference at zero temperature in the unbiased case, $\epsilon = 0$. In each panel the exact ML-MCTDH results from Ref \cite{zeroT} are the solid black dots, the dotted blue lines are direct MFT results, the solid red lines are FBTS results, and the solid blue lines are the results of our MF-GQME approach.}
\label{fig:5}
\end{figure}

In contrast to the spin-boson model at finite temperature, the zero-temperature limit is considered to be more challenging, as nuclear quantum effects in the bath become more prominent; a classical treatment of the bath distribution would predict only a single allowed bath initial configuration. In all the results shown here, the Wigner sampling of the bath ensures that zero-point energy is included exactly in the initial condition, although this is not guaranteed to be preserved by the approximate evolution prescribed by MFT or FBTS. In Fig. \ref{fig:5} we compare to exact multi-layer multi-configurational time-dependent Hartree (ML-MCTDH) results at zero-temperature in the nonadiabatic regime \cite{zeroT}.  As seen in Fig. \ref{fig:2} and Fig. \ref{fig:3} increasing the nonadibaticity, $\omega_c/\Delta$, or system-bath coupling, $\xi$, degrades the performance of MFT and FBTS at progressively shorter times although due to the lack of energetic bias the long time limits are in, probably fortuitously, good agreement. Suprisingly, FBTS performs worse than MFT in these regimes and again exhibits over-damped behavior resulting in slower relaxation. The MF-GQME results are in excellent agreement, with some very mild underdamping present in the case where the system-bath coupling is strong (middle panel). 

\begin{figure} 
\includegraphics[width=0.8\columnwidth,angle=0]{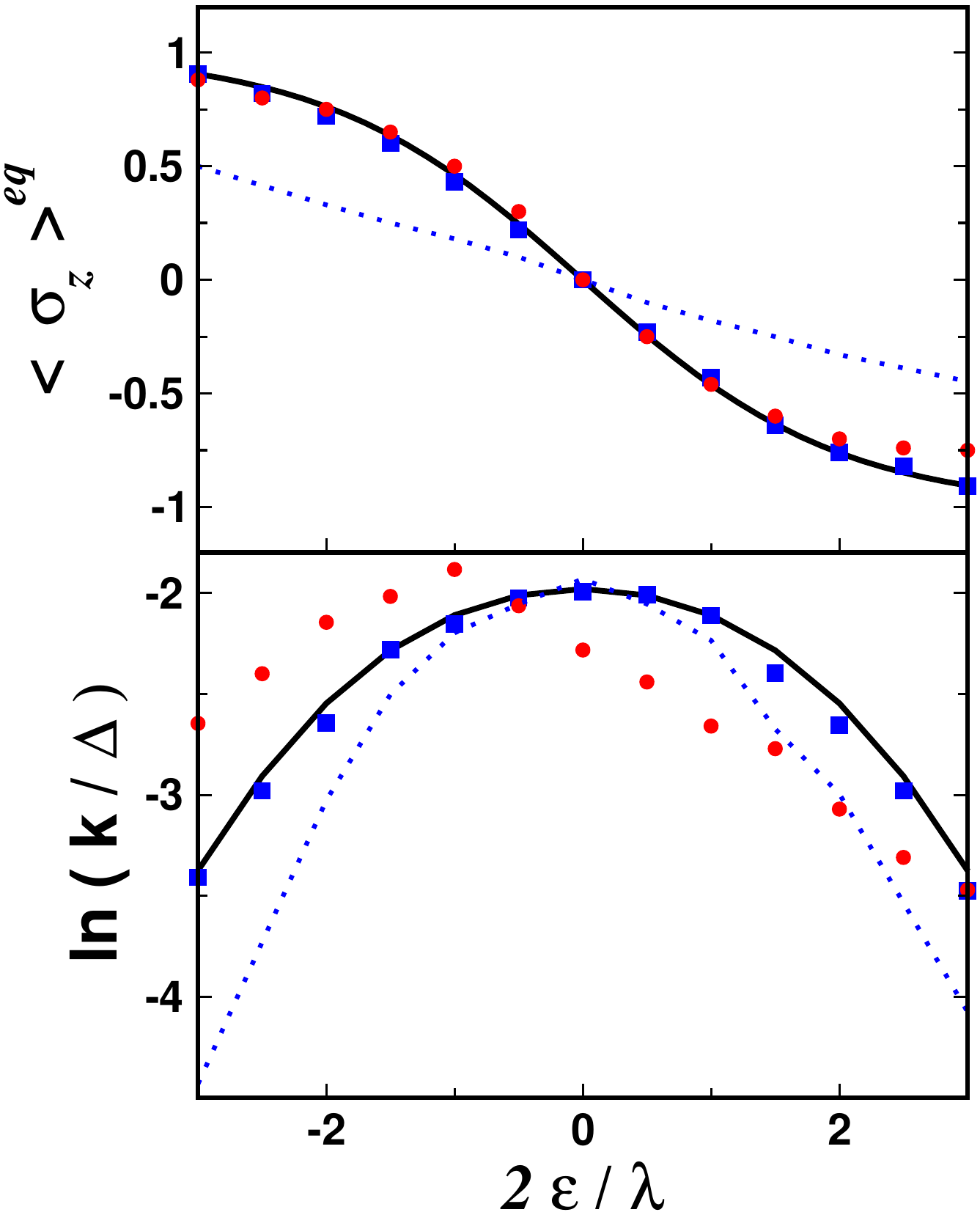} 
\caption{Marcus electron transfer regime, with $\omega_c = 10\Delta$, $\xi = 1.0$, $\beta\Delta  = 0.05$, and $\lambda = 2 \xi \omega_c$ is the reorganization energy. The top panel shows the equilibrium subsystem population difference after relaxation has occurred while the bottom panel shows the electron transfer rate as a function of driving force obtained from exponential fits to the population decay. In the top panel the solid black line is the Boltzmann distribution, in the bottom panel the solid black line is the Marcus rate, the dotted blue lines are MFT results, the solid red circles are FBTS results, and the solid blue squares are the results of our MF-GQME approach. }
\label{fig:6}
\end{figure}

In addition to the significant increase in accuracy afforded by MF-GQME over FBTS and direct MFT, this is accompanied by a significant increase in efficiency. As in Fig. \ref{fig:4}, the memory kernel decay times in Fig. \ref{fig:5} required to obtain the converged population dynamics are substantially shorter than the population decay times they generate, with memory kernel of lengths $\tau\Delta = 0.2$, $0.2$ and $0.1$ required to obtain the results in the top, middle and bottom panels respectively. This again reflects that the memory kernel becomes shorter as the system becomes more non-adiabatic. Due to the much faster decay of the memory kernel than the populations dynamics (which occurs in approximately $~5\Delta^{-1}$), the MF-GQME results in Fig. 4 are between 25 and 50 times cheaper to generate than direct MFT dynamics and up to 500 times cheaper than FBTS.

One of the most well-known difficulties of mean field theory occurs in the Marcus electron transfer regime \cite{marcus}. While most approximate dynamics approaches, like FSSH and MFT, are capable of qualitatively capturing the famous rate turnover as a function of driving force \cite{afssh,shi13,mjgqme,shs}, as shown in the bottom panel of Fig. \ref{fig:6}, MFT fails to correctly describe the donor-acceptor product ratios for the process. Much like the PBME approach  \cite{mjgqme}, the FBTS method qualitatively captures the rate turnover but has a slightly asymmetric shape. Also, perhaps surprisingly given their mean field treatment of the system-bath interactions, both PBME and FBTS give the correct equilibrium distribution at long times. The ability of FBTS to qualitatively and semiquantitatively capture Marcus turnover is consistent with recent PLDM results for a similar model of Marcus electron transfer that was simulated using a flux-side correlation function approach  \cite{pldm2}. The agreement between FBTS and PLDM is expected as the methods are identical when the subsystem Hamiltonian is chosen to be traceless \cite{fbts2}. They thus offer a similar tradeoff between cost and accuracy in the semiclassical hierarchy.

While FBTS performs well by including some dynamical correlations between the system and bath, the complete neglect of these correlations in direct MFT leads to poor performance. This is strongly pronounced in biased regimes (those with large electronic driving forces) where the reaction rates are underestimated and the long time population difference is too small. The MF-GQME approach resolves these issues, giving rise to quantitative agreement with the Marcus prediction for the rates and the Boltzmann distribution of the long time populations. In order to obtain the MF-GQME results in this regime $\tau\Delta  = 0.5$ time units of data were typically included in the memory kernels, and the population decay occurs on a timescale of a few hundred time units. This allowed for a $\approx200$ fold efficiency gain compared to direct MFT, in addition to a substantially improved accuracy.   

\section{Conclusions}\label{sec:conc}
Here we have shown that utilizing MFT within the GQME framework gives rise to nonadiabatic relaxation dynamics that are highly accurate across a wide range of physical regimes. This is accompanied by a computational savings of one or two orders of magnitude compared with direct MFT calculations, and up to three orders of magnitude over FBTS. This success can be rationalized based on the fact that the subsystem Liouville operator is treated exactly within the GQME and only the short-lived memory kernel term is approximated by MFT, which is accurate at short-times. The complementary nature of the combination of MFT with the GQME is amplified as the dynamics become more strongly nonadiabatic, as the memory kernel becomes increasingly short-lived compared to the subsystem population relaxation time. While MFT can fail at long-times when used directly, the timescale separation between the population dynamics and the memory kernel decay in these cases ensures that the dynamics can still be captured accurately using the MF-GQME. The MF-GQME approach is therefore expected to be efficient enough be used to study nonequilibrium relaxation problems in complex condensed phase systems at a very low computational cost, and it can be applied to any form of the system, bath or coupling between them, i.e. it is in no way limited to the linear coupling or harmonic bath invoked in the spin-boson model studied here.

Although MF-GQME offers probably the highest possible accuracy for the lowest possible cost, other similar calculations can be carried out using other dynamics methods such as FSSH, linearized and partially linearized approaches, and higher tier methods \cite{tully3,lscivr1,pbme,fbts1,pldm}. Indeed, at the opposite end of the hierarchy, very computationally demanding approaches such as the momentum-jump solution to the QCLE \cite{mj} can be made tractable when combined within the GQME framework  \cite{mjgqme}, greatly expanding its regime of applicability. In addition, the method for generating the memory kernel from system-dependent bath correlation functions adopted here represents just one possible way obtaining $\mathcal{K}$ from semiclassical trajectory-based simulation methods, and future work will explore these issues.

\section{Acknowledgements}
The authors would like to thank Tim Berkelbach, Will Pfalzgraff and Hans Andersen for helpful comments and a thorough reading of this manuscript. T.E.M. acknowledges support from a Terman fellowship, an Alfred P. Sloan Research fellowship, a Hellman Faculty Scholar Fund fellowship and Stanford University start-up funds. A.K. acknowledges a postdoctoral fellowship from the Stanford Center for Molecular Analysis and Design. N.B. was supported by the National Science Foundation Graduate Research Fellowship Program under Grant No. DGE-114747.

\section{Appendix}
\subsection{Explicit expressions for computing $\mathcal{K}_1$ and $\mathcal{K}_3$ for the spin-boson model}

According to the procedure outlined in Sec. (\ref{mf_eval}), the integration in  Eq. (25) is carried out over the phase space of the bath by Monte Carlo sampling initial conditions from $\left[\hat{\rho}_b^{eq} \mathcal{A}\right]_W(X,0)$ and generating dynamical trajectories to evaluate $ \mathcal{B}_W(X,\tau)$. 

If ${\mathcal{A}}$ is independent of momentum and linear in the coordinates of the bath, the Wigner transform of the operator product is \cite{imre67}
\begin{align}
[ \mathcal{A} \den]^*_W = [ \den \mathcal{A} ]_W=\mathcal{A}_W \rho_{b,W}^{eq} +\frac{i}{2 \hbar} \frac{\partial \mathcal{A}_W}{\partial R} \cdot \frac{\partial \rho_{b,W}^{eq}}{\partial P},
\end{align} which can be evaluated analytically for the harmonic bath employed in this study. 

In the spin-boson model, $\hat{\Lambda}=-\sum_j c_j \hat{R}_j$ and it's Wigner transform is $\Lambda_W = - \sum_j c_j R_j$. The Wigner transform of the equilibrium density for the isolated bath is \begin{multline}
\denb (X)= \prod_j \frac{\tanh{\beta \omega_j /2}}{\pi} \\
\times \text{exp} \left[ -\frac{2 \tanh{\beta \omega_j /2}}{\omega_j} \left( \frac{P^2}{2M_j}+\frac{M_j\omega_j^2}{2}R_j^2\right)\right].
\end{multline} 

In practice, our initial conditions for the bath degrees of freedom were sampled from the initial bath density (by taking out a factor of the bath density from Eq. (28)) and the trajectories were then re-weighted using the remaining term in Eq. (28).

The subsystem operator in the system-bath coupling part of the Hamiltonian is $\hat{S}=\hat{\sigma}_z$, and the matrix elements of $\hat{S}$ in the subsystem basis are given by
\begin{align}
\bra{m} \hat{S} \ket{n}&=\bra{m}   \hat{\sigma}_z \ket{n}=S_n \delta_{mn}  \label{Hsb}=(-1)^{n+1}\delta_{mn}.
\end{align}
where, $m$ and $n$ are eigenstates of the isolated subsystem, i.e. $\hat{\sigma}_z \ket{1}=\ket{1}$ and $\hat{\sigma}_z \ket{2}=-\ket{2}$. 

The elements of $\ko$ and $\kth$ are then given by
\widetext \begin{align}
(\ko)_{\alpha \alpha' \beta \beta'}(\tau) \quad &=& ((-1)^{\beta +\alpha}+ (-1)^{\beta+\alpha'+1}) \int dX \left[ \Lambda_W \rho_{b,W}^{eq} -\frac{i}{2 \hbar} \frac{\partial \Lambda_W}{\partial R} \cdot \frac{\partial \rho_{b,W}^{eq}}{\partial P} \right] (X,0) \Lambda_W(X,\tau) \rho_s^{\beta \beta'} (0) \rho_s^{\alpha' \alpha} (\tau) \nonumber \\ &&
+((-1)^{\beta'+\alpha+1}+(-1)^{\beta'+\alpha'}) \int dX \left[ \Lambda_W \rho_{b,W}^{eq} +\frac{i}{2 \hbar} \frac{\partial \Lambda_W}{\partial R} \cdot \frac{\partial \rho_{b,W}^{eq}}{\partial P} \right] (X,0) \Lambda_W(X,\tau) \rho_s^{\beta \beta'} (0) \rho_s^{\alpha' \alpha} (\tau),
 \label{eq:K1}
\end{align}
and 
\begin{align}
(\kth)_{\alpha \alpha ' \beta \beta '}  (\tau)\quad & = &(-1)^{\beta+1} \nonumber  
\int dX \left[ \Lambda_W \rho_{b,W}^{eq} -\frac{i}{2 \hbar} \frac{\partial \Lambda_W}{\partial R} \cdot \frac{\partial \rho_{b,W}^{eq}}{\partial P} \right] (X,0) \rho_s^{\beta \beta'} (0) \rho_s^{\alpha' \alpha} (\tau) \nonumber \\ && +(-1)^{\beta'}  \int dX \left[ \Lambda_W \rho_{b,W}^{eq} +\frac{i}{2 \hbar} \frac{\partial \Lambda_W}{\partial R} \cdot \frac{\partial \rho_{b,W}^{eq}}{\partial P} \right] (X,0) \rho_s^{\beta \beta'} (0) \rho_s^{\alpha' \alpha} (\tau), 
 \label{eq:K3}
\end{align} where $\rho_s^{\alpha \alpha'} = c_{\alpha} c^{*}_{\alpha'}$ is the subsystem RDM, and $c_{\alpha}$ and $c_{\alpha'}$ are the coefficients of the subsystem wavefunction. 


\begin{thebibliography}{58}
\expandafter\ifx\csname natexlab\endcsname\relax\def\natexlab#1{#1}\fi
\expandafter\ifx\csname bibnamefont\endcsname\relax
  \def\bibnamefont#1{#1}\fi
\expandafter\ifx\csname bibfnamefont\endcsname\relax
  \def\bibfnamefont#1{#1}\fi
\expandafter\ifx\csname citenamefont\endcsname\relax
  \def \citenamefont#1{#1}\fi
\expandafter\ifx\csname url\endcsname\relax
  \def\url#1{\texttt{#1}}\fi
\expandafter\ifx\csname urlprefix\endcsname\relax\def\urlprefix{URL }\fi
\providecommand{\bibinfo}[2]{#2}
\providecommand{\eprint}[2][]{\url{#2}}

\bibitem[{ \citenamefont{Mclachlan}(1964)}]{mft}
\bibinfo{author}{\bibfnamefont{A.~D.}~\bibnamefont{McLachlan}}
  \bibinfo{journal}{Mol. Phys.} \textbf{\bibinfo{volume}{8}},
  \bibinfo{pages}{39} (\bibinfo{year}{1964}).

   \bibitem[{ \citenamefont{Sun et~al.}(1998) \citenamefont{Sun, Wang, and
  Miller}}]{lscivr1}
\bibinfo{author}{\bibfnamefont{X.}~\bibnamefont{Sun}},
  \bibinfo{author}{\bibfnamefont{H.~B.} \bibnamefont{Wang}}, \bibnamefont{and}
  \bibinfo{author}{\bibfnamefont{W.~H.} \bibnamefont{Miller}},
  \bibinfo{journal}{J. Chem. Phys.} \textbf{\bibinfo{volume}{109}},
  \bibinfo{pages}{7064} (\bibinfo{year}{1998}).

\bibitem[{ \citenamefont{Shi and Geva}(2003)}]{lscivr2}
\bibinfo{author}{\bibfnamefont{Q.}~\bibnamefont{Shi}} \bibnamefont{and}
  \bibinfo{author}{\bibfnamefont{E.}~\bibnamefont{Geva}}, \bibinfo{journal}{J.
  Chem. Phys.} \textbf{\bibinfo{volume}{118}}, \bibinfo{pages}{8173}
  (\bibinfo{year}{2003}).

\bibitem[{ \citenamefont{Poulsen et al}(2003)}]{lscivr3}
\bibinfo{author}{\bibfnamefont{J.~A.}~\bibnamefont{Poulsen}},
\bibinfo{author}{\bibfnamefont{G.}~\bibnamefont{Nyman}}, \bibnamefont{and}
  \bibinfo{author}{\bibfnamefont{P.~J.} \bibnamefont{Rossky}},
  \bibinfo{journal}{J. Chem. Phys.} \textbf{\bibinfo{volume}{119}},
  \bibinfo{pages}{12179} (\bibinfo{year}{2003}).

\bibitem[{ \citenamefont{Kim et~al.}(2008) \citenamefont{Kim, Nassimi, and
  Kapral.}}]{pbme}
\bibinfo{author}{\bibfnamefont{H.}~\bibnamefont{Kim}},
  \bibinfo{author}{\bibfnamefont{A.}~\bibnamefont{Nassimi}}, \bibnamefont{and}
  \bibinfo{author}{\bibfnamefont{R.}~\bibnamefont{Kapral.}},
  \bibinfo{journal}{J. Chem. Phys.} \textbf{\bibinfo{volume}{129}},
  \bibinfo{pages}{084102} (\bibinfo{year}{2008}).

 \bibitem[{ \citenamefont{Bonella and Coker}(2005)}]{landmap}
\bibinfo{author}{\bibfnamefont{S.}~\bibnamefont{Bonella}} \bibnamefont{and}
  \bibinfo{author}{\bibfnamefont{D.~F.} \bibnamefont{Coker}},
  \bibinfo{journal}{J. Chem. Phys.} \textbf{\bibinfo{volume}{122}},
  \bibinfo{pages}{194102} (\bibinfo{year}{2005}).

 \bibitem[{ \citenamefont{Huo and Coker}(2011)}]{pldm}
\bibinfo{author}{\bibfnamefont{P.}~\bibnamefont{Huo}} \bibnamefont{and}
  \bibinfo{author}{\bibfnamefont{D.~F.} \bibnamefont{Coker}},
  \bibinfo{journal}{J. Chem. Phys.} \textbf{\bibinfo{volume}{135}},
  \bibinfo{pages}{201101} (\bibinfo{year}{2011}).
  
 \bibitem[{ \citenamefont{Hsieh and Kapral}(2008)}]{fbts1}
\bibinfo{author}{\bibfnamefont{C.-Y.}~\bibnamefont{Hsieh}} \bibnamefont{and}
  \bibinfo{author}{\bibfnamefont{R.}~\bibnamefont{Kapral.}},
  \bibinfo{journal}{J. Chem. Phys.} \textbf{\bibinfo{volume}{137}},
  \bibinfo{pages}{22A507} (\bibinfo{year}{2012}).

 \bibitem[{ \citenamefont{Hsieh and Kapral}(2008)}]{fbts2}
\bibinfo{author}{\bibfnamefont{C.-Y.}~\bibnamefont{Hsieh}} \bibnamefont{and}
  \bibinfo{author}{\bibfnamefont{R.}~\bibnamefont{Kapral.}},
  \bibinfo{journal}{J. Chem. Phys.} \textbf{\bibinfo{volume}{138}},
  \bibinfo{pages}{134110} (\bibinfo{year}{2013}).
 
 \bibitem[{ \citenamefont{MacKernan et~al.}(2002) \citenamefont{MacKernan,
  Ciccotti, and Kapral.}}]{sstp}
\bibinfo{author}{\bibfnamefont{D.}~\bibnamefont{MacKernan}},
  \bibinfo{author}{\bibfnamefont{G.}~\bibnamefont{Ciccotti}}, \bibnamefont{and}
  \bibinfo{author}{\bibfnamefont{R.}~\bibnamefont{Kapral.}},
  \bibinfo{journal}{J. Phys.: Condens. Matter} \textbf{\bibinfo{volume}{14}},
  \bibinfo{pages}{9069} (\bibinfo{year}{2002}).

 \bibitem[{ \citenamefont{MacKernan et~al.}(2008) \citenamefont{MacKernan,
  Ciccotti, and Kapral.}}]{mj}
\bibinfo{author}{\bibfnamefont{D.}~\bibnamefont{MacKernan}},
  \bibinfo{author}{\bibfnamefont{R.}~\bibnamefont{Kapral.}},\bibnamefont{and}
  \bibinfo{author}{\bibfnamefont{G.}~\bibnamefont{Ciccotti}}, 
  \bibinfo{journal}{J. Phys. Chem. B} \textbf{\bibinfo{volume}{112}},
  \bibinfo{pages}{424} (\bibinfo{year}{2008}).

   \bibitem[{ \citenamefont{Dunkel et~al.}(2008) \citenamefont{Dunkel, Bonella, and
  Coker}}]{ildm}
\bibinfo{author}{\bibfnamefont{E.}~\bibnamefont{Dunkel}},
  \bibinfo{author}{\bibfnamefont{S.}~\bibnamefont{Bonella}}, \bibnamefont{and}
  \bibinfo{author}{\bibfnamefont{D.~F.} \bibnamefont{Coker}},
  \bibinfo{journal}{J. Chem. Phys.} \textbf{\bibinfo{volume}{129}},
  \bibinfo{pages}{114106} (\bibinfo{year}{2008}).

   \bibitem[{ \citenamefont{Tully and Preston}(1971)}]{tully1}
\bibinfo{author}{\bibfnamefont{J.~C.} \bibnamefont{Tully}},\bibnamefont{and}
\bibinfo{author}{\bibfnamefont{R.~K.} \bibnamefont{Preston}},
  \bibinfo{journal}{J. Chem. Phys.} \textbf{\bibinfo{volume}{55}},
  \bibinfo{pages}{562} (\bibinfo{year}{1971}).

 \bibitem[{ \citenamefont{Tully}(1990)}]{tully2}
\bibinfo{author}{\bibfnamefont{J.~C.} \bibnamefont{Tully}},
  \bibinfo{journal}{J. Chem. Phys.} \textbf{\bibinfo{volume}{93}},
  \bibinfo{pages}{1061} (\bibinfo{year}{1990}).

\bibitem[{\citenamefont{Bittner and Rossky}(1995)}]{bittner_rossky}
\bibinfo{author}{\bibfnamefont{E.~R.} \bibnamefont{Bittner}},\bibnamefont{and}
\bibinfo{author}{\bibfnamefont{P.~J.} \bibnamefont{Rossky}},
  \bibinfo{journal}{J. Chem. Phys.} \textbf{\bibinfo{volume}{103}},
  \bibinfo{pages}{8130} (\bibinfo{year}{1995}).

\bibitem[{ \citenamefont{Landry and Subotnik}(2012)}]{afssh}
  \bibinfo{author}{\bibfnamefont{B.~R.}~\bibnamefont{Landry}}, \bibnamefont{and}
\bibinfo{author}{\bibfnamefont{J.~E.} \bibnamefont{Subotnik}} 
  \bibinfo{journal}{J. Chem. Phys.} \textbf{\bibinfo{volume}{137}},
  \bibinfo{pages}{22A513} (\bibinfo{year}{2012}).

\bibitem[{ \citenamefont{Landry and Subotnik}(2012)}]{tully3}
  \bibinfo{author}{\bibfnamefont{B.~R.}~\bibnamefont{Landry}}, 
 \bibinfo{author}{\bibfnamefont{M.~J.}~\bibnamefont{Falk}}, \bibnamefont{and}
\bibinfo{author}{\bibfnamefont{J.~E.} \bibnamefont{Subotnik}} 
  \bibinfo{journal}{J. Chem. Phys.} \textbf{\bibinfo{volume}{139}},
  \bibinfo{pages}{211101} (\bibinfo{year}{2013}).

 \bibitem[{ \citenamefont{Miller}(2001)}]{miller}
\bibinfo{author}{\bibfnamefont{W.~H.} \bibnamefont{Miller}},
  \bibinfo{journal}{J. Phys. Chem. A} \textbf{\bibinfo{volume}{105}},
  \bibinfo{pages}{2942} (\bibinfo{year}{2001}).

\bibitem[{ \citenamefont{Thoss and Wang}(2004)}]{thoss_wang}
  \bibinfo{author}{\bibfnamefont{M.}~\bibnamefont{Thoss}}, \bibnamefont{and}
\bibinfo{author}{\bibfnamefont{H.} \bibnamefont{Wang}} 
  \bibinfo{journal}{Ann. Rev. Phys. Chem.} \textbf{\bibinfo{volume}{55}},
  \bibinfo{pages}{299} (\bibinfo{year}{2004}).

 \bibitem[{ \citenamefont{Kapral}(2006)}]{kapral06}
\bibinfo{author}{\bibfnamefont{R.} \bibnamefont{Kapral}},
  \bibinfo{journal}{Annu. Rev. Phys. Chem.} \textbf{\bibinfo{volume}{57}},
  \bibinfo{pages}{1239} (\bibinfo{year}{2006}).

 \bibitem[{ \citenamefont{Nakajima}(1958)}]{gqme1}
\bibinfo{author}{\bibfnamefont{S.} \bibnamefont{Nakajima}},
  \bibinfo{journal}{Prog. Theor. Phys.} \textbf{\bibinfo{volume}{20}},
  \bibinfo{pages}{948} (\bibinfo{year}{1958}).

\bibitem[{ \citenamefont{Zwanzig}(1960)}]{gqme2}
\bibinfo{author}{\bibfnamefont{R.} \bibnamefont{Zwanzig}},
  \bibinfo{journal}{J. Chem. Phys.} \textbf{\bibinfo{volume}{33}},
  \bibinfo{pages}{1338} (\bibinfo{year}{1960}).
  
   \bibitem[{ \citenamefont{Shi and Geva}(2004)}]{sng2}
\bibinfo{author}{\bibfnamefont{Q.}~\bibnamefont{Shi}} \bibnamefont{and}
  \bibinfo{author}{\bibfnamefont{E.}~\bibnamefont{Geva}}, \bibinfo{journal}{J.
  Chem. Phys.} \textbf{\bibinfo{volume}{120}}, \bibinfo{pages}{10647}
  (\bibinfo{year}{2004}).
  
 \bibitem[{ \citenamefont{Shi and Geva}(2004)}]{sng3}
\bibinfo{author}{\bibfnamefont{Q.}~\bibnamefont{Shi}} \bibnamefont{and}
  \bibinfo{author}{\bibfnamefont{E.}~\bibnamefont{Geva}}, \bibinfo{journal}{J.
  Chem. Phys.} \textbf{\bibinfo{volume}{121}}, \bibinfo{pages}{3393}
  (\bibinfo{year}{2004}).  
  
   \bibitem[{ \citenamefont{Kelly and Markland}(2013)}]{mjgqme}
\bibinfo{author}{\bibfnamefont{A.} \bibnamefont{Kelly}} \bibnamefont{and}
\bibinfo{author}{\bibfnamefont{T.~E.} \bibnamefont{Markland}},  
  \bibinfo{journal}{J. Chem. Phys.} \textbf{\bibinfo{volume}{139}},
    \bibinfo{pages}{014104} (\bibinfo{year}{2013}).

 \bibitem[{ \citenamefont{Kapral and Ciccotti}(1999)}]{qcle}
\bibinfo{author}{\bibfnamefont{R.}~\bibnamefont{Kapral}} \bibnamefont{and}
  \bibinfo{author}{\bibfnamefont{G.}~\bibnamefont{Ciccotti}},
  \bibinfo{journal}{J. Chem. Phys.} \textbf{\bibinfo{volume}{110}},
  \bibinfo{pages}{8919} (\bibinfo{year}{1999}).
  
  \bibitem[{\citenamefont{Grunwald et~al.}(2009)\citenamefont{Grunwald, Kelly,
  and Kapral}}]{chapter}
\bibinfo{author}{\bibfnamefont{R.}~\bibnamefont{Grunwald}},
  \bibinfo{author}{\bibfnamefont{A.}~\bibnamefont{Kelly}}, \bibnamefont{and}
  \bibinfo{author}{\bibfnamefont{R.}~\bibnamefont{Kapral}}, in
  \emph{\bibinfo{booktitle}{Energy Transfer Dynamics in Biomaterial Systems}},
  edited by \bibinfo{editor}{\bibfnamefont{I.}~\bibnamefont{Burghardt}}
  (\bibinfo{publisher}{Springer}, \bibinfo{address}{Berlin},
  \bibinfo{year}{2009}), pp. \bibinfo{pages}{383--413}.

\bibitem[{ \citenamefont{Zhang et al}(2006)}]{geva06}
\bibinfo{author}{\bibfnamefont{M.-L.}~\bibnamefont{Zhang}},
\bibinfo{author}{\bibfnamefont{B. J.}~\bibnamefont{Ka}} \bibnamefont{and}
  \bibinfo{author}{\bibfnamefont{E.}~\bibnamefont{Geva}}, \bibinfo{journal}{J.
  Chem. Phys.} \textbf{\bibinfo{volume}{125}}, \bibinfo{pages}{044106}
  (\bibinfo{year}{2006}).

  \bibitem[{ \citenamefont{Shi and Geva}(2003)}]{sng1}
\bibinfo{author}{\bibfnamefont{Q.}~\bibnamefont{Shi}} \bibnamefont{and}
  \bibinfo{author}{\bibfnamefont{E.}~\bibnamefont{Geva}}, \bibinfo{journal}{J.
  Chem. Phys.} \textbf{\bibinfo{volume}{119}}, \bibinfo{pages}{12063}
  (\bibinfo{year}{2003}).

\bibitem[{ \citenamefont{Muhlbacher and Rabani}(2003)}]{muhlbacher08}
\bibinfo{author}{\bibfnamefont{L.} \bibnamefont{M\"uhlbacher}}, \bibnamefont{and}
\bibinfo{author}{\bibfnamefont{E.} \bibnamefont{Rabani}},
  \bibinfo{journal}{Phys. Rev. Lett.} \textbf{\bibinfo{volume}{100}},
    \bibinfo{pages}{176403} (\bibinfo{year}{2008}).
    
\bibitem[{ \citenamefont{Cohen and Rabani}(2011)}]{cohen11}
\bibinfo{author}{\bibfnamefont{G.}~\bibnamefont{Cohen}} \bibnamefont{and}
  \bibinfo{author}{\bibfnamefont{E.}~\bibnamefont{Rabani}}, \bibinfo{journal}{Phys. Rev. B} 
  \textbf{\bibinfo{volume}{84}}, \bibinfo{pages}{075150}
  (\bibinfo{year}{2011}).
  
\bibitem[{ \citenamefont{Wilner and Rabani}(2014)}]{wilner}
\bibinfo{author}{\bibfnamefont{E.~Y.}~\bibnamefont{Wilner}},
\bibinfo{author}{\bibfnamefont{H.}~\bibnamefont{Wang}}, 
\bibinfo{author}{\bibfnamefont{M.}~\bibnamefont{Thoss}},\bibnamefont{and}
  \bibinfo{author}{\bibfnamefont{E.}~\bibnamefont{Rabani}}, \bibinfo{journal}{Phys. Rev. B} 
  \textbf{\bibinfo{volume}{90}}, \bibinfo{pages}{115145}
  (\bibinfo{year}{2014}).

 \bibitem[{ \citenamefont{Redfield}(1957}]{redfield}
\bibinfo{author}{\bibfnamefont{A.~G.}~\bibnamefont{Redfield}}, \bibinfo{journal}{IBM
 J. Res. Dev.} \textbf{\bibinfo{volume}{1}}, \bibinfo{pages}{19}
  (\bibinfo{year}{1957}).

\bibitem[{ \citenamefont{Leggett}(1987)}]{spinboson1}
\bibinfo{author}{\bibfnamefont{A.} \bibnamefont{Leggett}},
\bibinfo{author}{\bibfnamefont{S.} \bibnamefont{Chakravarty}},
\bibinfo{author}{\bibfnamefont{A.} \bibnamefont{Dorsey}},
\bibinfo{author}{\bibfnamefont{M.} \bibnamefont{Fisher}},
\bibinfo{author}{\bibfnamefont{A.} \bibnamefont{Garg}},  \bibnamefont{and}
\bibinfo{author}{\bibfnamefont{R.} \bibnamefont{Zwerger}}, 
  \bibinfo{journal}{Rev. Mod. Phys.} \textbf{\bibinfo{volume}{59}},
  \bibinfo{pages}{1} (\bibinfo{year}{1987}).

 \bibitem[{ \citenamefont{Weiss}(1992)}]{spinboson2}
\bibinfo{author}{\bibfnamefont{U.} \bibnamefont{Weiss}},
  \emph{\bibinfo{title}{Quantum Dissipative Systems}} (\bibinfo{publisher}{World Scientific}, \bibinfo{address}{Singapore}, \bibinfo{year}{1992}).

   \bibitem[{ \citenamefont{Tao and Miller}(2010)}]{taomiller}
\bibinfo{author}{\bibfnamefont{G.} \bibnamefont{Tao}} \bibnamefont{and}
\bibinfo{author}{\bibfnamefont{W.~H.} \bibnamefont{Miller}},  
  \bibinfo{journal}{J. Phys. Chem. Lett.} \textbf{\bibinfo{volume}{1}},
    \bibinfo{pages}{891} (\bibinfo{year}{2010}).

   \bibitem[{ \citenamefont{Kelly and Rhee}(2011)}]{kellyrhee}
\bibinfo{author}{\bibfnamefont{A.} \bibnamefont{Kelly}} \bibnamefont{and}
\bibinfo{author}{\bibfnamefont{Y.~M.} \bibnamefont{Rhee}},  
  \bibinfo{journal}{J. Phys. Chem. Lett.} \textbf{\bibinfo{volume}{2}},
    \bibinfo{pages}{808} (\bibinfo{year}{2011}).

\bibitem[{ \citenamefont{Berkelbach et. al}(2012)}]{rdmh1}
\bibinfo{author}{\bibfnamefont{T.~C.} \bibnamefont{Berkelbach}},  
\bibinfo{author}{\bibfnamefont{D.~R.} \bibnamefont{Reichman}},\bibnamefont{and}
\bibinfo{author}{\bibfnamefont{T.~E.} \bibnamefont{Markland}},  
  \bibinfo{journal}{J. Chem. Phys.} \textbf{\bibinfo{volume}{136}},
    \bibinfo{pages}{034113} (\bibinfo{year}{2012}).

\bibitem[{ \citenamefont{Berkelbach et. al}(2012)}]{rdmh2}
\bibinfo{author}{\bibfnamefont{T.~C.} \bibnamefont{Berkelbach}},  
\bibinfo{author}{\bibfnamefont{T.~E.} \bibnamefont{Markland}},\bibnamefont{and}
\bibinfo{author}{\bibfnamefont{D.~R.} \bibnamefont{Reichman}}, 
  \bibinfo{journal}{J. Chem. Phys.} \textbf{\bibinfo{volume}{136}},
    \bibinfo{pages}{084104} (\bibinfo{year}{2012}).

 \bibitem[{ \citenamefont{Huo and Coker}(2012)}]{pldm1}
\bibinfo{author}{\bibfnamefont{P.}~\bibnamefont{Huo}} \bibnamefont{and}
  \bibinfo{author}{\bibfnamefont{D.~F.} \bibnamefont{Coker}},
  \bibinfo{journal}{J. Chem. Phys.} \textbf{\bibinfo{volume}{137}},
  \bibinfo{pages}{22A535} (\bibinfo{year}{2012}).

 \bibitem[{ \citenamefont{Huo and Coker}(2011)}]{ildm2}
\bibinfo{author}{\bibfnamefont{P.}~\bibnamefont{Huo}} \bibnamefont{and}
  \bibinfo{author}{\bibfnamefont{D.~F.} \bibnamefont{Coker}},
  \bibinfo{journal}{J. Chem. Phys.} \textbf{\bibinfo{volume}{133}},
  \bibinfo{pages}{184108} (\bibinfo{year}{2011}).

\bibitem[{ \citenamefont{Makrarov and Makri}(1994)}]{quapi1}
\bibinfo{author}{\bibfnamefont{D.~E.} \bibnamefont{Makarov}}, \bibnamefont{and}
\bibinfo{author}{\bibfnamefont{N.} \bibnamefont{Makri}},
\bibinfo{journal}{Chem. Phys. Lett.} \textbf{\bibinfo{volume}{221}},
\bibinfo{pages}{482} (\bibinfo{year}{1994}).

\bibitem[{ \citenamefont{Makri and Makrarov}(1994)}]{quapi2}
\bibinfo{author}{\bibfnamefont{N.} \bibnamefont{Makri}}, \bibnamefont{and}
\bibinfo{author}{\bibfnamefont{D.~E.} \bibnamefont{Makarov}},
\bibinfo{journal}{J. Chem. Phys.} \textbf{\bibinfo{volume}{102}},
\bibinfo{pages}{4600} (\bibinfo{year}{1994}).

\bibitem[{ \citenamefont{Makri and Makrarov}(1994)}]{quapi3}
\bibinfo{author}{\bibfnamefont{N.} \bibnamefont{Makri}}, \bibnamefont{and}
\bibinfo{author}{\bibfnamefont{D.~E.} \bibnamefont{Makarov}},
\bibinfo{journal}{J. Chem. Phys.} \textbf{\bibinfo{volume}{102}},
 \bibinfo{pages}{4611} (\bibinfo{year}{1994}).

\bibitem[{ \citenamefont{Wang and Thoss}(2008)}]{zeroT}
\bibinfo{author}{\bibfnamefont{H.} \bibnamefont{Wang}}, \bibnamefont{and}
\bibinfo{author}{\bibfnamefont{M.} \bibnamefont{Thoss}},
\bibinfo{journal}{New J. Phys.} \textbf{\bibinfo{volume}{10}},
 \bibinfo{pages}{115005} (\bibinfo{year}{2008}).

\bibitem[{ \citenamefont{Marcus}(1964)}]{marcus}
\bibinfo{author}{\bibfnamefont{R.A.} \bibnamefont{Marcus}}
  \bibinfo{journal}{Annu. Rev. Phys. Chem.} \textbf{\bibinfo{volume}{15}},
  \bibinfo{pages}{155} (\bibinfo{year}{1964}).

  \bibitem[{ \citenamefont{Shi}(2013)}]{shi13}
\bibinfo{author}{\bibfnamefont{W.} \bibnamefont{Xie}},
\bibinfo{author}{\bibfnamefont{S.} \bibnamefont{Bai}},
\bibinfo{author}{\bibfnamefont{L.} \bibnamefont{Zhu}}, \bibnamefont{and}
\bibinfo{author}{\bibfnamefont{Q.} \bibnamefont{Shi}},
  \bibinfo{journal}{J. Phys. Chem. A},
  \bibinfo{pages}{jp400462f} (\bibinfo{year}{2013}).
      
\bibitem[{ \citenamefont{Schwerdtfeger et. al}(2014)}]{shs}
\bibinfo{author}{\bibfnamefont{C.~A.} \bibnamefont{Schwerdtfeger}},  
\bibinfo{author}{\bibfnamefont{A.~V.} \bibnamefont{Soudackov}},\bibnamefont{and}
\bibinfo{author}{\bibfnamefont{S.} \bibnamefont{Hammes-Schiffer}},  
  \bibinfo{journal}{J. Chem. Phys.} \textbf{\bibinfo{volume}{140}},
    \bibinfo{pages}{034113} (\bibinfo{year}{2012}).    
  
        \bibitem[{ \citenamefont{Huo, Miller and Coker}(2011)}]{pldm2}
\bibinfo{author}{\bibfnamefont{P.}~\bibnamefont{Huo}},
\bibinfo{author}{\bibfnamefont{T.~F.}~\bibnamefont{Miller III}} \bibnamefont{and}
  \bibinfo{author}{\bibfnamefont{D.~F.} \bibnamefont{Coker}},
  \bibinfo{journal}{J. Chem. Phys.} \textbf{\bibinfo{volume}{139}},
  \bibinfo{pages}{151103} (\bibinfo{year}{2012}).
      
    \bibitem[{ \citenamefont{Imre et~al.}(1967) \citenamefont{Imre, \"{O}zizmir,
  Rosenbaum, and Zwiefel}}]{imre67}
\bibinfo{author}{\bibfnamefont{K.}~\bibnamefont{Imre}},
  \bibinfo{author}{\bibfnamefont{E.}~\bibnamefont{\"{O}zizmir}},
  \bibinfo{author}{\bibfnamefont{M.}~\bibnamefont{Rosenbaum}},
  \bibnamefont{and} \bibinfo{author}{\bibfnamefont{P.~F.}
  \bibnamefont{Zwiefel}}, \bibinfo{journal}{J. Math. Phys.}
  \textbf{\bibinfo{volume}{5}}, \bibinfo{pages}{1097} (\bibinfo{year}{1967}).
  
\end{thebibliography}
\end{document}